\documentclass{ws-ijmpd}

\begin{document}

\markboth{S. Schramm and V. Dexheimer} {Compact Stars in Hadron
and Quark-Hadron Models}

%
\catchline{}{}{}{}{}
%

\title{
Compact Stars in Hadron and Quark-Hadron Models }

\author{Stefan Schramm}

\address{FIAS, Goethe University, Ruth Moufang Str. 1\\
60438 Frankfurt, Germany
\\
schramm@th.physik.uni-frankfurt.de}

\author{Ver\^onica Antocheviz Dexheimer}

\address{Department of Physics, Gettysburg College, 300 North Washington Street\\
Gettysburg, PA 17325, USA }
\maketitle

\begin{history}
\received{Day Month Year}
\revised{Day Month Year}
\comby{Managing Editor}
\end{history}

\begin{abstract}
We investigate strongly interacting dense matter and neutron
stars using a flavor-SU(3) approach based on a non-linear
realization of chiral symmetry as well as a hadronic
flavor-SU(2) parity-doublet model. We study chiral symmetry
restoration and the equation of state of stellar matter and
determine neutron star properties using different sets of
degrees of freedom. Finally, we include quarks in the model
approach. We show the resulting phase diagram as well as hybrid
star solutions for this model.
\end{abstract}


\section{Introduction}	

The study of strongly interacting matter at extreme conditions
of temperature and density is at the forefront of modern
nuclear physics. This regime covers the physics of
ultrarelativistic heavy-ion collisions as well as important
aspects of nuclear astrophysics. Whereas the determination of
the phase structure of excited strongly interacting matter at
high temperature and baryon densities, as pursued in heavy-ion
collisions, involves relatively high temperatures, the
complementary study of the structure of compact stars is
directly related to the properties of very dense and rather
cold matter.

In order to study all these regimes in a unified model approach
we developed an effective chiral SU(3) model that can be
studied over the whole relevant range of chemical potentials
and temperatures. Reflecting the correct degrees of freedom of
the system after the transition to a quark-gluon plasma state
we include quarks and the Polyakov loop as order parameter for
the deconfinement phase transition.

\section{The Hadronic Model}

The model used in our calculations is an extended
$\sigma-\omega$ chiral model, based on a non-linear realization
of chiral symmetry, that includes the lowest SU(3) multiplets
of baryons and mesons as well as an effective field that mimics
the QCD scale anomaly. The expectation values of the scalar
isoscalar fields correspond to the non-strange isoscalar
($\sigma$ field), isovector ($\delta$) and strange
($s\overline{s}$) chiral quark condensates ($\zeta$).

The interactions between baryons and the scalar and vector
mesons are linear and read (assuming static systems, including
only the time component of the vector fields)
\begin{equation}
\label{L_BM+V}
{\cal L}_{\rm BM} =
-\sum_{i}   \overline{\psi}_i \left( g_{i\sigma}\sigma + g_{i\delta}\delta + g_{i\zeta}\zeta
\right) \psi_i
\end{equation}
\begin{equation} {\cal L}_{\rm BV} = -\sum_{i} \overline{\psi}_i
\left( g_{i\omega}\gamma_0\omega^0 + g_{i\rho}\gamma_0\rho^0 + g_{i \phi}\gamma_0 \phi^0
\right) \psi_i ~,
\end{equation}
where $\omega, \rho,$ and $\phi$ are the non-strange isoscalar,
isovector and strange vector fields, respectively. The index
$i$ sums over the baryon octet ($N$, $\Lambda$, $\Sigma$,
$\Xi$). Additional terms include mass terms and quartic
self-interactions of the vector mesons, whereas the
self-interactions of the scalar mesons induce the spontaneous
breaking of chiral symmetry. The effect of non-zero current
quark masses is modelled by introducing an explicit
chiral-symmetry breaking term (for more details on the full
Lagrangian see \cite{chiral2}). The vacuum masses of the
baryons are generated through their coupling to the chiral
condensates.

For the baryon-vector couplings $g_{i\omega}$ and $g_{i\phi}$
pure $f$-type coupling is assumed as discussed in
\cite{chiral2}, $g_{i\omega} = (n^i_q-n^i_{\bar{q}})
g_{8}^V$\,, $g_{i\phi}   = -(n^i_s-n^i_{\bar{s}}) \sqrt{2}
g_{8}^V$\,,
where 
$g_8^V$ denotes the vector coupling of the baryon octet and
$n^i$ the number of constituent quarks of species $i$ in a
given hadron. The resulting relative couplings are in
accordance with additive quark model constraints, coupling only
the $\phi$ vector mesons to strange baryons.

All parameters of the model are fixed by either symmetry
relations, hadronic vacuum observables or nuclear matter
saturation properties (see \cite{chiral2}). In addition, the
model also provides a satisfactory description of realistic
(finite-size and isospin asymmetric) nuclei and neutron stars
\cite{chiral2,apj}.
Using this approach the equations of motion are solved for
isospin symmetric nuclear matter as well as star matter by
including leptons and requiring charge neutrality. Fig. 1 shows
the resulting equation of state of cold symmetric nuclear and
star matter. The parameters were slightly tuned from the fit
values used in calculations of finite nuclei \cite{deformed} to
explore the range of  possible maximum star masses, while
retaining a reasonable description of saturated nuclear matter.
The nuclear compressibility has a still acceptable value of 297
MeV \cite{apj}. The results for three choices of baryonic
degrees of freedom are shown, which comprise (a) nucleons, (b)
nucleons and hyperons, and (c) nucleons, hyperons and spin 3/2
baryonic resonances. As expected, with increasing number of
particle species the equation of state becomes softer at larger
densities.
\begin{figure}[pb]
\centerline{\psfig{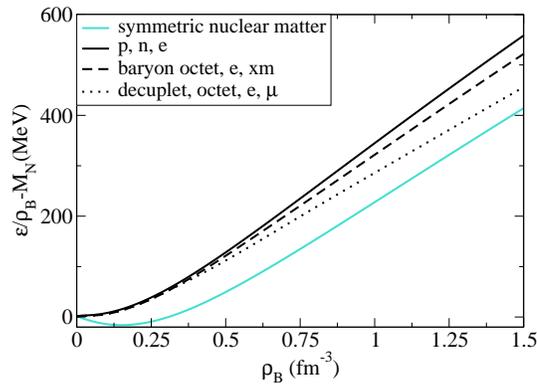}}
\vspace*{8pt}
\caption{Equation of State of isospin-symmetric nuclear matter and stellar matter including nucleons, nucleons and hyperons, and in addition spin3/2 baryons.}
\end{figure}
Using these equations of state in a calculation of the
properties of a spherical and static neutron star, integrating
the Tolman-Oppenheimer-Volkoff (TOV) equations \cite{apj}, we
obtain maximum star masses between 1.93 and 2.12 solar masses
depending on the number of degrees of freedom taken into
account (Fig. 2). One can observe that even including $\Delta$
baryons in the calculation rather large star masses with
typical radii of about 12km are possible.
\begin{figure}[pb]
\centerline{\psfig{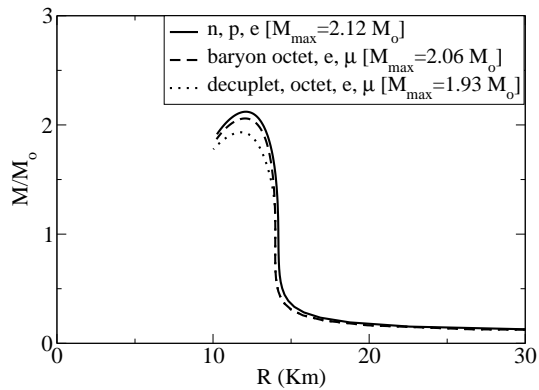}}
\vspace*{8pt}
\caption{Mass-Radius diagram for neutron star solutions using different EOS as shown in Fig. 1.}
\end{figure}
Looking into the corresponding particle densities, shown in
Fig. 3, the hyperon content of the star is relatively small
with $\Lambda$ baryons appearing at about 3 times nuclear
matter groundstate density and the $\Sigma^-$ at 4 $\rho_0$,
respectively. The latter result is due to the fact that in the
model the $\Sigma$ potential in nuclear matter is positive
($U_\Sigma = 5.35 $MeV).
\begin{figure}[pb]
\centerline{\psfig{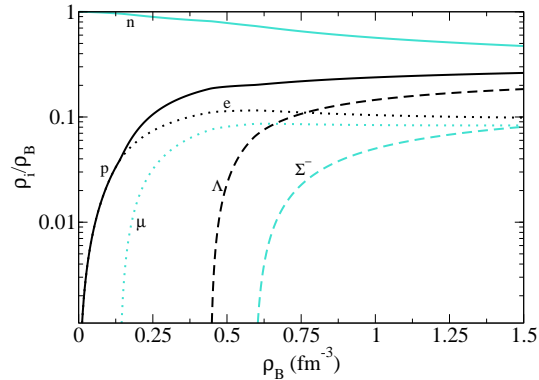}}
\vspace*{8pt}
\caption{Particle abundances as function of density for a star including nucleons and hyperons.}
\end{figure}
In case one includes the baryonic spin 3/2 decuplet, that is
$\Delta$ baryons for the densities occurring in compact stars,
the particle mix looks very different as can be seen in Fig. 4.
The $\Delta^-$ is populated above 2 $\rho_0$ and the hyperons
are shifted to larger densities. Note, however, that the
results depend strongly on the largely unknown value of the
vector coupling of the $\Delta$ to the $\omega$ meson, here
taken to be identical to the nucleon-$\omega$ coupling.
\begin{figure}[pb]
\centerline{\psfig{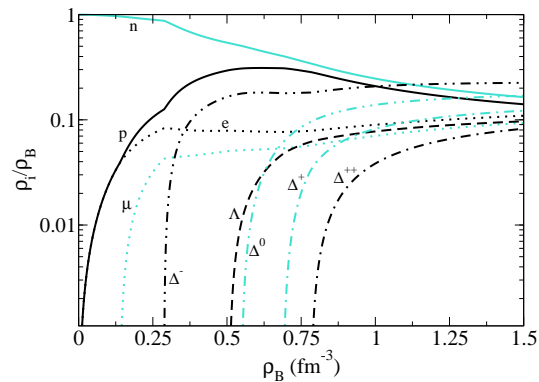}}
\vspace*{8pt}
\caption{Particle abundances as function of density for a star including nucleons, hyperons, and $\Delta$ resonances.}
\end{figure}

One possible effect that might occur in neutron stars is meson,
in particular $K^-$, condensation \cite{kaplan}. Using our
model, in a calculation of the critical densities for the onset
of the condensation in neutron and proto-neutron stars we find
that for cold stars the condensate sets in at values of about 6
times nuclear matter saturation density whereas for
proto-neutron stars the value is even higher. These values are
beyond the maximum central densities inside of the stars such
that kaon condensation does not affect stellar properties
\cite{kaon1,kaon2}.

\section{The Parity-Doublet Model}
Using an alternative approach of formulating a chirally
symmetric Lagrangian we consider the parity doublet model
originally discussed in \cite{deTar}. Here, one adopts the
so-called ``mirror assignment'' for the positive and negative
parity nucleon states ($N_+$ and $N_-$, we restrict ourselves
to SU(2), in this case), grouping both states in the same
multiplet. Under left- and right-handed transformations
$SU_L(2) \times SU(2)_R$ transformations the two nucleon fields
$\psi_1$ and $\psi_2$ transform non-trivially as:
\begin{eqnarray}
\psi_{1R} \longrightarrow R \psi_{1R} \  & , \hspace{1cm} &   \psi_{1L}
\longrightarrow L \psi_{1L} \ , \label{mirdef1} \\
\psi_{2R} \longrightarrow L \psi_{2R} \  & , \hspace{1cm} &   \psi_{2L}
\longrightarrow R \psi_{2L} \ . \label{mirdef2}
\end{eqnarray}
In a similar way as $\sigma$ and $\pi$ are parity partners in
the linear realization of chiral symmetry this allows for a
chirally invariant nucleonic mass term in the Lagrangian:
\begin{eqnarray}
&&m_{0}( \bar{\psi}_2 \gamma_{5} \psi_1 - \bar{\psi}_1
      \gamma_{5} \psi_2 ) =  m_0 (\bar{\psi}_{2L} \psi_{1R} -
        \bar{\psi}_{2R} \psi_{1L} - \bar{\psi}_{1L} \psi_{2R} +
        \bar{\psi}_{1R} \psi_{2L}) \ , \label{chinvmass}
\end{eqnarray}
where $m_0$ represents a bare mass parameter. After
diagonalizing the quadratic terms in the Lagrangian one obtains
the unmixed fields, $N_+$ and $N_-$, that are the standard
nucleon and its parity partner. In the limit of chiral symmetry
restoration ($\sigma = 0$) both nucleonic states are
degenerate, but attain a finite mass $m_0$. This doubling of
degenerate nucleonic states is the characteristic feature of
this type of chiral model.

Using the parity-doublet ansatz combined with a linear sigma
model including the $\omega$ meson (for achieving the correct
nuclear matter saturation properties) and the $\rho$ meson (for
reproducing the phenomenological value for the asymmetry energy
of 32 MeV) one can determine the corresponding field values and
the equation of state by solving the equations of motion of the
model in mean-field approximation \cite{parity,parityrha}. In
Fig. 5 the resulting $\sigma$ field, the scalar condensate, is
shown. The field becomes smaller with increasing density
tending towards chiral symmetry restoration. The structure in
the curve for star matter at about 2 times $\rho_0$ corresponds
to the onset of populating the parity partner of the neutron.
This can directly been seen in Fig. 6. Above 4 $\rho_0$ the
parity partner of the proton shows up.

Inserting the equation of state into the star calculation one
obtains a maximum star mass of 1.85 solar masses. Including
sizeable vector meson self-interaction terms leads to very low
star masses in contradiction to values of observed stars. Going
beyond the mean-field approximation we repeated the
calculations using the Relativistic Hartree Approximation
(RHA), which in general generates comparable results slightly
lowering the nuclear matter compressibility \cite{parityrha}.

In general, for the parity partners to be populated in the
interior of the star a relatively low vacuum mass of the $N^-$
baryon below 1380 MeV is required. In addition, in order to
maintain reasonably small values for the compressibility one
has to assume a high value of the mass parameter $m_0 > 800
$MeV. A full analysis of the question whether with these values
one can still be in accordance with low-energy hadronic data
like pion-nucleon scattering is still an open question
\cite{parityrha}.

\begin{figure}[pb]
\centerline{\psfig{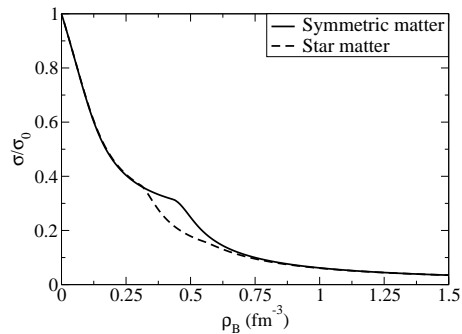}}
\vspace*{8pt}
\caption{Scalar condensate as function of density in the parity-doublet model. The mass for the
parity partner $N_-$ is set to 1200 MeV.}
\end{figure}

\begin{figure}[pb]
\centerline{\psfig{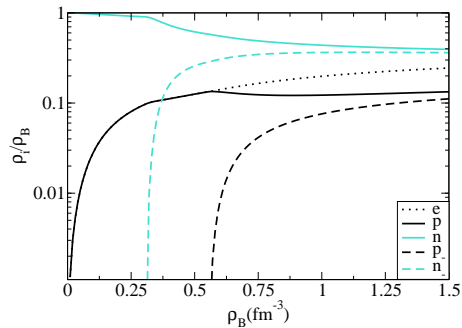}}
\vspace*{8pt}
\caption{Densities of particle species as function of density in the parity-doublet model.}
\end{figure}

\section{Including Quarks}

Lattice QCD calculations show that the phase transition from
hadrons to quarks and gluons is a smooth crossover at vanishing
chemical potential. Some lattice calculations at finite
chemical potential $\mu_B$, which are still notoriously
difficult to perform, suggest a critical end-point of a line of
first-order phase transitions in the T-$\mu_B$ plane
\cite{fodor}. Calculations that connect hadrons and quarks in
two different model approaches are not able to reproduce such a
transition behavior. Therefore we extend our hadronic model by
including quarks and an effective field $\Phi$ for the Polyakov
loop, following \cite{PNJL,Ratti1,Ratti2}, in order to describe
the deconfinement phase transition in a single unified approach
\cite{quarks}. The potential for the Polyakov loop reads:
\begin{equation}
U=(a_0T^4+a_1\mu_B^4+a_2T^2\mu_B^2)\Phi^2 + a_3T_0^4\ln{(1-6\Phi^2+8\Phi^3-3\Phi^4)}.
\end{equation}

\begin{figure}[pb]
\centerline{\psfig{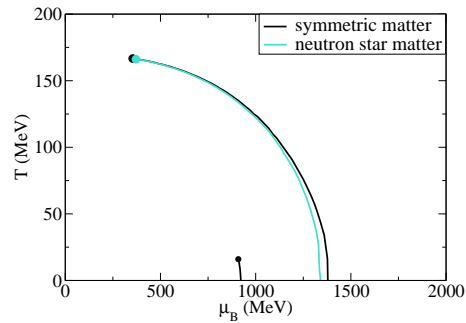}}
\vspace*{8pt}
\caption{Phase diagram in $\mu_B$ and $T$ for the Hadron-Quark model. First-order transition lines for the
chiral and deconfinement transitions as well as for the liquid-gas phase transitions are shown as calculated in the model.}
\end{figure}

The parameters are obtained by fitting the values to lattice
QCD results at vanishing chemical potential. The
$\mu_B$-dependent terms are fixed by reproducing the critical
end-point as computed in \cite{fodor}. The switch between quark
and baryon degrees of freedom is achieved in an effective way,
shifting the baryon/quark masses to high/small values for
small/high Polyakov loop values using a simple additional mass
term for quarks and baryons, such that their effective masses
read:
\begin{eqnarray}
&m_{b}^*=g_{b\sigma}\sigma+g_{b\delta}\tau_3\delta+g_{b\zeta}\zeta+m_{0b}+g_{b\Phi} \Phi^2,&
\end{eqnarray}
\begin{eqnarray}
&m_{q}^*=g_{q\sigma}\sigma+g_{q\delta}\tau_3\delta+g_{q\zeta}\zeta+ m_{0q}+g_{q\Phi}(1-\Phi)~~,&
\end{eqnarray}
where $m_0$ are small explicit mass terms and $g_{b\Phi}$ and
$g_{q\Phi}$ are the coupling constants of the Polyakov loop to
baryons and quarks, respectively \cite{quarks}. Using this
approach we obtain the phase diagram shown in Fig. 7. Note that
we also obtain a realistic first-order liquid-gas phase
transition. The transition to quarks at zero temperature occurs
at $4 \rho_0$.

It is straightforward to calculate star masses in this model.
The result in Fig. 8 shows a maximum star mass of 1.92 solar
masses. Using a Maxwell construction, assuming local charge
neutrality of the star matter, the family of stable stars ends
with the appearance of quarks and a significant softening of
the equation of state. Assuming global charge neutrality one
obtains a mixed phase of quarks and hadrons in the inner 2km
core of the maximum mass star.

\begin{figure}[pb]
\centerline{\psfig{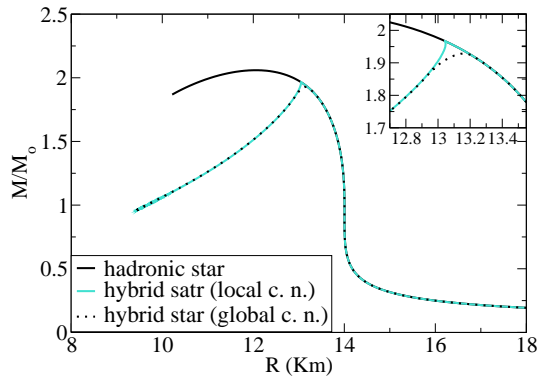}}
\vspace*{8pt}
\caption{Mass-radius relation of compact stars in the Hadron-Quark model. The inset shows the different results for a Maxwell and Gibbs
construction of the phase transition. The Gibbs construction generates a mixed hadron-quark phase that smooths the sharp transition.}
\end{figure}


\section{References}

\begin{thebibliography}{00}
\bibitem{chiral2}
  P.~Papazoglou, D.~Zschiesche, S.~Schramm, J.~Schaffner-Bielich, H.~St\"ocker and W.~Greiner,
  Phys.\ Rev.\  C {\bf 59}, 411 (1999).
\bibitem{apj}
  V.~Dexheimer and S.~Schramm,
  Astrophys.\ J.\ {\bf 683}, 943 (2008).
\bibitem{deformed} S. Schramm, Phys.
    Rev. C66 064310 (2002).
\bibitem{kaplan}
    D. B. Kaplan and A. E. Nelson, Phys. Lett. B {\bf 175}, 57 (1986);
    A. E. Nelson and D. B. Kaplan, {\it ibid}, 192, 193 (1987).
\bibitem{kaon1} A. Mishra, A. Kumar, S. Sanyal, S. Schramm,
    Eur.Phys.J.A41:205-213 (2009).
\bibitem{kaon2} A. Mishra, A. Kumar, S. Sanyal, V. Dexheimer,
    submitted to Phys. Rev. C., arxiv 0905.3518 [nucl-th].
\bibitem{deTar}
 C. DeTar, T. Kunihiro, Phys. Rev. D {\bf 39}, 2805 (1989).
\bibitem{parity} V. Dexheimer, S. Schramm, and D. Zschiesche,
    Phys.
    Rev. C 77, 025803 (2008).

\bibitem{parityrha} V. Dexheimer, G. Pagliara, L. Tolos, J.
    Schaffner-Bielich, S. Schramm, Eur. Phys. J. A 38, 105-113 (2008)
\bibitem{fodor}
  Z.~Fodor and S.~D.~Katz, JHEP {\bf 0404} (2004) 050.


\bibitem{PNJL}
  K.~Fukushima,
  Phys.\ Lett.\  B {\bf 591}, 277 (2004)
  [arXiv:hep-ph/0310121].

\bibitem{Ratti1}
  C.~Ratti, M.~A.~Thaler and W.~Weise,
  Phys.\ Rev.\  D {\bf 73}, 014019 (2006)

\bibitem{Ratti2}
  S.~Roessner, C.~Ratti and W.~Weise,
  Phys.\ Rev.\  D {\bf 75}, 034007 (2007)
\bibitem{quarks} V. Dexheimer and S. Schramm,
    , preprint arxiv:0901.1748 [astro-ph], submitted to Phys Rev C;
    V. Dexheimer and S. Schramm, Nucl. Phys.
    A827,  988 (2009).








\end{thebibliography}
\end{document}